\documentclass[prb,twocolumn,showpacs,amsmath,amssymb,preprintnumbers,superscriptaddress]{revtex4}
\usepackage{dcolumn}
\usepackage{bm}
\usepackage{graphicx}
\usepackage{color}

\begin{document}

\title{Large exchange bias and low temperature glassy state in frustrated triangular-lattice antiferromagnet Ba$_3$NiIr$_2$O$_9$}

\author{Shobha Gondh}\affiliation{School of Physical Sciences, Jawaharlal Nehru University, New Delhi - 110067, India.}
\author{Manju Mishra Patidar}\affiliation{UGC-DAE Consortium for Scientific Research, Indore - 452001, India.}
\author{Kranti Kumar}\affiliation{UGC-DAE Consortium for Scientific Research, Indore - 452001, India.}
\author{M. P. Saravanan}\affiliation{UGC-DAE Consortium for Scientific Research, Indore - 452001, India.}
\author{V. Ganesan}\affiliation{UGC-DAE Consortium for Scientific Research, Indore - 452001, India.}
\author{A. K. Pramanik}\email{akpramanik@mail.jnu.ac.in}\affiliation{School of Physical Sciences, Jawaharlal Nehru University, New Delhi - 110067, India.}

\begin{abstract}
Here, we report both ac and dc magnetization, thermodynamic and electric properties of hexagonal Ba$_3$NiIr$_2$O$_9$. The Ni$^{2+}$ (spin-1) forms layered triangular-lattice and interacts antiferromagnetically while Ir$^{5+}$ is believed to act as magnetic link between the layers. This complex magnetic interaction results in magnetic frustration leading to a spin-glass transition at $T_f$ $\sim$ 8.5 K. The observed magnetic relaxation and aging effect also confirms the nonequilibrium ground state. The system further shows large exchange bias which is tunable with cooling field. Below the Curie-Weiss temperature $\theta_{CW}$ ($\sim$ -29 K), the magnetic specific heat $C_m$ displays a broad hump and at low temperature follows $C_m = \gamma T^\alpha$ dependence where both $\gamma$ and $\alpha$ show dependence on temperature and magnetic field. A sign change in magnetoresistace is observed which is due to an interplay among magnetic moment, field and spin-orbit coupling.       
\end{abstract}

\pacs{75.47.Lx, 75.40.Cx, 75.50.Lk, 73.43.Qt}

\maketitle
\section{Introduction}
Magnetic materials that combine an antiferromagnetic (AFM) interaction and triangular or tetrahedral lattice structure harbor geometrical frustration which often gives rise to many unusual magnetic properties; the examples include spin liquids, spin glasses, spin ices, etc.\cite{Ramirez,Nakatsuji1,Zhou,Cheng,Okamoto,Nag,Nakatsuji,MacLaughlin,Krizan,Fu,Greedan,Raju,Ma,Gaudet} While both the magnetic frustration and the quantum spin fluctuations play key role here, long time back it was predicted by Anderson that materials with such high degree of ground state degeneracy can not lead to long-range magnetic ordering based on only nearest-neighbor interactions.\cite{Anderson} However, presence of auxiliary interactions in terms of nearest neighbor or interlayer interactions help to lift the degeneracy which often induces various exotic magnetic states.\cite{Nag,Fu,Cheng,Villain}

Here, we report an experimental investigation of complex interaction driven spin glass (SG) behavior in 3$d$-5$d$ based layered Ba$_3$NiIr$_2$O$_9$. This material belongs to the class of triangular-lattice antiferromagnet (TLAF) where the layered triangular-lattice of Ni$^{2+}$ (3$d^2$, $S$ = 1) interact antiferromagnetically which are connected by dimers of Ir$^{5+}$ (5$d^4$). While for Ni$^{2+}$ the usual $e_g$ and $t_{2g}$ splitting due to crystal field effect describes well the electronic structure but in case of 5$d$ Ir$^{5+}$, the strong spin-orbit coupling (SOC) further splits the $t_{2g}$ state into low lying $J_{eff}$ = 3/2 quartet and $J_{eff}$ = 1/2 doublet.\cite{Kim, Kim1} Following this $J_{eff}$ model, therefore a fully filled $J_{eff}$ = 3/2 is expected for Ir$^{5+}$ which gives it a nonmagnetic character ($J_{eff}$ = 0). However, this SOC driven $J_{eff}$ states is under debate where a breakdown of $J_{eff}$ picture under strong noncubic crystal field in distorted IrO$_6$ octahedra has been shown to attach magnetic moment for Ir$^{5+}$ ions.\cite{Nag,Bremholm,Bhowal,Kharkwal,Cao,Khan} On quantitative point, using neutron diffraction experiment recently we have shown an on-site ordered moment for Ir$^{5+}$ is around 0.5 $\mu_B$/site in double-perovskite material Sr$_2$FeIrO$_6$.\cite{Kharkwal} Therefore, the present Ba$_3$NiIr$_2$O$_9$ is not a purely 2-dimensional magnetic system while though magnetic Ni$^{2+}$ ions form triangular-lattice in layer. We believe that in opposed to conventional $J_{eff}$ model the Ir$^{5+}$ ions possess moment, hence the dimers of Ir$^{5+}$ act as weak magnetic link between the Ni layers, giving it a quasi 2-dimensional magnetic interaction. The Ir$^{5+}$ induced SOC and interlayer magnetic exchange in combination with dominant intralayer AFM interaction will render complex magnetic exchange which would impact its magnetic and electronic transport properties significantly.

Unlike its isostructural spin-1/2 system i.e., (Ba,Sr)$_3$CuSb$_2$O$_9$) which shows quantum spin-liquid (QSL) behavior,\cite{Zhou, Kundu} the Ni$^{2+}$ based spin-1 systems such as, Ba$_3$NiSb$_2$O$_9$, $A_3$NiNb$_2$O$_9$ ($A$ = Ba, Sr, and Ca) usually exhibit long-range AFM ordering.\cite{Cheng,Lu,Hwang} Though the magnetic lattice as well as magnetic interaction favors magnetic frustration, a reduced quantum spin fluctuation because of high spin probably does not allow QSL state in these materials. Exception is, however, high-pressure synthesized Ba$_3$NiSb$_2$O$_9$ demonstrating the QSL behavior\cite{Cheng} while a controversy exists between QSL and SG for NiGa$_2$S$_4$.\cite{Nakatsuji, MacLaughlin} Our experimental results comprising of both dc and ac magnetization and specific heat suggest a highly degenerate ground state in Ba$_3$NiIr$_2$O$_9$ which is characterized with a spin-glass (SG) behavior. Due to this complex spin interactions, the system shows large exchange bias tunable with cooling fields. A crossover between positive and negative magnetoresistance has been observed which is due to an interplay between magnetic moment and SOC.                   
                             
\section{Experimental Methods}														
Polycrystalline sample Ba$_3$NiIr$_2$O$_9$ (BNIO) has been prepared by solid state reaction method. The starting ingredient materials are taken as BaCO$_3$, NiO and IrO$_2$ in stoichiometric ratio with phase purity of $>$ 99.99\% (Sigma-Aldrich). These mixture has been ground well and heated to 800$^\circ$ C for 24 h in powder form for calcination. The powders are then pressed into pellets and given heat treatment at 1050$^\circ$ C and 1070$^\circ$ C for 24 h each with an intermediate grinding. Each time the heating and cooling rate has been fixed to 3$^\circ$ C/min. To check phase purity and determine the crystal structure, powder x-ray diffraction (XRD) measurement has been done using a diffractometer by Rigaku (model: Miniﬂex600) with CuK$_\alpha$ radiation at room temperature. The XRD data have been collected in $2\theta$ range 10 to 90$^0$ with a step size 0.02$^\circ$. The collected XRD data are analyzed with Rietveld refinement using FULLPROF program.\cite{suite} The both ac and dc magnetization, specific heat and electronic transport are measured with SQUID and PPMS by Quantum Design.

The XRD data along with Rietveld refinement are presented in Fig. 1(a) showing a very good fitting. We have obtained the refinement parameters $R_p$ = 16.9, $R_{wp}$ = 20.1 and $R_{exp}$ = 14.2 giving the `goodness of fit (GOF)' parameter $R_{wp}$/$R_{exp}$ $\sim$ 1.42 which together with low $\chi^2$ $\sim$ 2  suggest good refinement. The refinement shows a single phase Ba$_3$NiIr$_2$O$_9$ which crystallizes into hexagonal structure with \textit{P6$_3$/mmc} symmetry (space group 194). The obtained lattice parameters $a$ and $c$ are 5.7542(3) {\AA} and 14.2713(8) {\AA}, respectively which gives $c/a$ ratio 2.48 and the unit cell volume 409.23(5) {\AA$^3$. The atomic positions and occupancy are given in Table I. Fig. 1(b) shows the structural unit of BNIO showing Ni layer and Ir dimers. The Ni ions form 2-dimensional (2D) layers of triangular-lattice in $ab$-plane with an interatomic distance 5.7542(3) {\AA} and interlayer separation 7.1375(4) {\AA} where the layers are connected by dimers of face-sharing Ir$_2$O$_9$ bioctahedra (Fig. 1(c)). 

\begin{table}
\caption{\label{label} The atomic positions and occupancies for different elements of BNIO (space group - \textit{P6$_3$/mmc}) obtained from Rietveld refinement.}
\begin{ruledtabular}
\begin{tabular}{ccccc}
Atom &Wyck. &x &y &z\\         
\hline
Ba1 &2b &0.00000 &0.00000 &0.25000\\
Ba2 &4f &0.33330 &0.66666 &0.91051(19)\\
Ni &2a &0.00000 &0.00000 &0.00000\\
Ir &4f &0.33333 &0.66666 &0.15403(12)\\
O1 &6h &0.49657(3) &0.99370(3) &0.25000\\
O2 &12k &0.164770(3) &0.32949(3) &0.41862(8)\\
\end{tabular}
\end{ruledtabular}
\end{table}

\begin{figure}
	\centering
		\includegraphics[width=8.5cm]{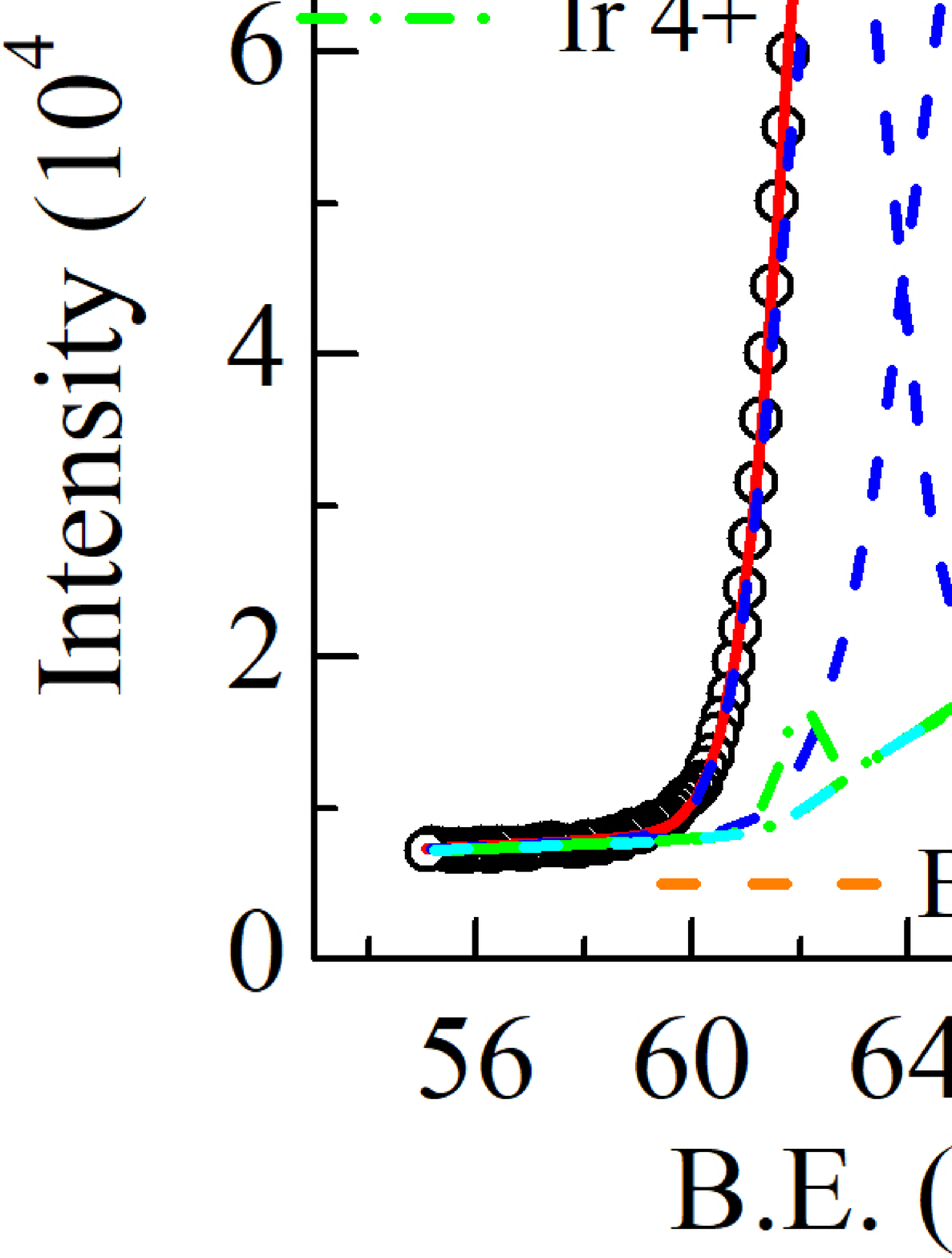}
	\caption{(a) shows room temperature XRD plot along with Rietveld refinement for Ba$_3$NiIr$_2$O$_9$. (b) shows the structural unit cell along with Ni and Ir atomic positions and (c) shows the triangular-lattice arrangement of Ni-atoms in layer of unit cell (top-view). The room temperature x-ray photoemission spectroscopy data are shown in (d) and (e) for Ba$_3$NiIr$_2$O$_9$. (d) shows the Ir-4$f$ core level spectra and (e) shows the Ni-2$p_{3/2}$ core level spectra along with fitting.}
	\label{fig:Fig1}
\end{figure}

Given that the charge states of constituent elements play an important role in physical properties of any system, we have done room temperature x-ray photoemission spectroscopy (XPS) measurements to understand the cationic charge states in Ba$_3$NiIr$_2$O$_9$. The Fig. 1(d) shows the Ir-4$f$ core level spectra along with fitting. The XPS data can be fitted well with two peaks corresponding to spin-orbit split Ir-4$f_{7/2}$ and Ir-4$f_{5/2}$ levels along with a satellite peak at binding energy (B.E.) 69.29 eV. From fitting, it is confirmed that dominant contribution comes from Ir$^{5+}$ state (dotted blue) which shows peaks at B.E. 62.67 and 65.67 eV, respectively with a separation of 3.0 eV. While this Ir$^{5+}$ is expected from its charge balance, we also find a small contribution of Ir$^{4+}$ state (dotted green) with the spin-orbit split peaks occurring at B.E. 62.0 and 65.2 eV, respectively with a energy separation of 3.2 eV.\cite{Kharkwal,Bhatti,Harish} Our analysis shows amount of Ir$^{5+}$ and Ir$^{4+}$ is around 98 and 2 \%, respectively. The presence of small amount Ir$^{4+}$ may be due to nonstiochiometry in the material, and this charge state coexistence has commonly been seen in different iridates.\cite{Kharkwal,Bhatti,Harish,Zhu,Liu,Nag} Fig. 1(e) shows Ni-2$p_{3/2}$ core level spectra along with fitting. The data has been fitted well with two peaks related to Ni$^{2+}$ with Ni-2$p_{3/2}$ peak at 854.79 eV (dotted blue) and its satellite occurring at 861.47 eV (dotted green).\cite{Babita,Elipe,Peck} The XPS analysis demonstrate a Ir$^{5+}$ and Ni$^{2+}$ ionic charge state in agreement with our expectation which imply that layers of magnetic Ni$^{2+}$ (spin-1) are separated by Ir$^{5+}$ where the later is believed to provide an interlayer magnetic exchange path, hence forming a complex interaction scenario at low temperature. Note, that a significant difference in ionic radii as well as in ionic states between Ni$^{2+}$ (0.69 {\AA}) and Ir$^{5+}$ (0.57 {\AA}) would minimize the intersite mixing of metal ions.

\section{Results and Discussion}
Magnetic susceptibility $\chi$ (=$M/H$) measured in 1 and 10 kOe magnetic field, following zero field cooled (ZFC) and field cooled (FC) protocol, are shown against temperature $T$ in Fig. 2(a). The $\chi (T)$ exhibit a sharp rise below 100 K with an onset of bifurcation around 65 K in 1 kOe field. At low-$T$, the $\chi_{ZFC}$ shows a prominent peak at $T_p$ = 6.7 K while the $\chi_{FC}$ shows a continuous increase down to 2 K, where both are typical to SG behavior. In 10 kOe, this bifurcation is, however, largely suppressed and the peak in $\chi_{ZFC}$ shifts to lower temperature at 6 K. A close look in low field $\chi_{ZFC}$ shows anomalies around 65, 28 and 13 K which is prominent in its derivative $d\chi/dT$ (upper inset  of Fig. 2(a)) showing a shoulder around 65 and 28 K (marked by down arrows) and a dip around 13 K. The $\chi_{ZFC}$ has further been measured in different fields (0.5, 1, 5, 10, 30 and 50 kOe) which show the anomaly in $\chi(T)$ becomes prominent in low field i.e., at 0.5 kOe (Fig. 2(b)). As seen in figure, with increasing fields the peak in $\chi_{ZFC}$ becomes broader, the $T_p$ shifts to lower temperature and $\chi$ decreases showing a similarity with SG behavior\cite{Luo, Anand, Kitaoka} while in case of long-range AFM transition the $\chi$ increases with measuring fields.\cite{Hwang} The variation of peak temperature $T_p$ is shown in Fig. 2(c) which shows a linear decrease of $T_p$ with an increasing magnetic field.

\begin{figure}
	\centering
		\includegraphics[width=8.5cm]{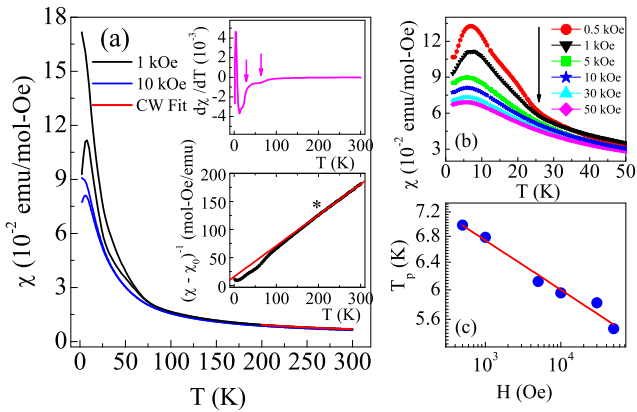}
	\caption{(a) shows the temperature dependence of dc susceptibility ($\chi$) following ZFC and FC measurements in 1 and 10 kOe for Ba$_3$NiIr$_2$O$_9$. The solid line (red) in high temperature is due to modified Curie-Weiss fitting. The upper inset shows temperature derivative $d\chi/dT$ where the vertically down arrows mark the temperature 65 and 28 K. The lower inset shows corrected inverse susceptibility ($\chi$ - $\chi_0$)$^{-1}$ vs $T$ plotting along with a linear fitting (red line) while the asterisk ($\ast$) represents the temperature for deviation from linearity. (b) shows the susceptibility $\chi$ plotted with temperature at different magnetic fields. (c) The magnetic field dependent peak temperature $T_p$ is shown along with linear fitting.}
	\label{fig:Fig2}
\end{figure}

The plotting of $\chi^{-1}(T)$ shows a deviation from linearity below 200 K (lower inset of Fig. 2(a)). The fitting of $\chi(T)$ with modified Curie-Weiss (CW) law, $\chi$ = $\chi_0$ + $C$/($T$-$\theta_{CW}$) (red line in Fig. 2(a)) gives temperature independent component $\chi_0$ = 1.3 $\times$ 10$^{-3}$ emu/mol-Oe, the Curie constant $C$ = 1.82 and the Curie-Weiss temperature $\theta_{CW}$ = -28.7 K. The obtained $\theta_{CW}$ is in order with other Ni based TLAF materials and imply a dominant AFM exchange between Ni$^{2+}$ ions. From $C$, an effective paramagnetic moment has been calculated to be $\mu_{eff}$ =3.81 $\mu_B$/f.u. which, however, comes out higher than the spin-only expected value 2.83 $\mu_B$/f.u. for $S$ = 1. In other way, the obtained $\mu_{eff}$ gives the Lande $g$-factor $g$ = 2.7 which appears to be higher than other Ni$^{2+}$ based systems where the $g$ value ranges between 2.2 - 2.5.\cite{Cheng, Hwang, Carlin, Shirata} This suggests that the Ir$^{5+}$, which is expected to be nonmagnetic ($J_{eff}$ = 0) following $J_{eff}$ model, contributes to the moment. This indication comes from our recent neutron diffraction measurements for Sr$_2$FeIrO$_6$ showing an ordered moment $\sim$ 0.5(3) $\mu_B$/site for Ir$^{5+}$ which also agrees with GGA+$U$+SOC based electronic structure calculation.\cite{Kharkwal} A state of nonmagnetic Ir$^{5+}$ has also been shown in other studies.\cite{Nag,Bhowal,Cao}

Given that Ir$_2$O$_9$ bioctahedra acts as linking path between Ni layers (Fig. 1(b)), this nonmagnetic Ir$^{5+}$ not only opens up an interaction between Ir-Ir within dimer but also promotes a weak interlayer magnetic coupling (i.e., Ir$^{5+}$-O-Ir$^{5+}$, Ir$^{5+}$-O-Ni$^{2+}$). We believe that interlayer magnetic exchange is of ferromagnetic (FM) type, and the multiple anomalies in low field $\chi_{ZFC}$ is likely due to the ordering of different exchange paths. For an estimate, if one assumes that each Ni$^{2+}$ ion interacts only with six nearest neighbor ions ($z$) in plane, then following Heisenberg model $J\sum_{<i,j>}S_i.S_j$ the exchange interaction $J$ can be calculated from $\theta_{CW}$ = $[-zJS(S+1)]/3k_B$ as $J/k_B$ = 7.2 K, which closely agrees with the $T_p$ value in Fig. 2(a). Given that Ni-based other TLAF systems i.e., Ba$_3$NiSb$_2$O$_9$ ($T_N$ = 13.5 K),\cite{Cheng} (Ba,Sr,Ca)$_3$NiNb$_2$O$_9$ ($T_N$ $\sim$ 5 K)\cite{Lu} exhibit long-range ordering at low temperatures, the onset of magnetic irreversibility in $\chi(T)$ around 65 K is unlikely due to an ordering of in-plane Ni$^{2+}$ spins. Instead, we believe that this irreversibility and anomalies in $\chi(T)$ around 65, 28 and 13 K are caused by short-range spin correlations, where a combination of dominating intralayer AFM and weak interlayer FM interactions causes glassy behavior at low temperature. Such a FM and AFM competition driven SG dynamics has been commonly observed in dilute alloys \cite{Villain} or even in pyrochlore Na$_3$Co(CO$_3$)$_2$Cl with a competing nearest-neighbor AFM and next-nearest-neighbor FM interactions.\cite{Fu} Nonetheless, the present BNIO presents an intriguing example of complex interaction driven SG like freezing at low temperature.    

\begin{figure}
	\centering
		\includegraphics[width=8.5cm]{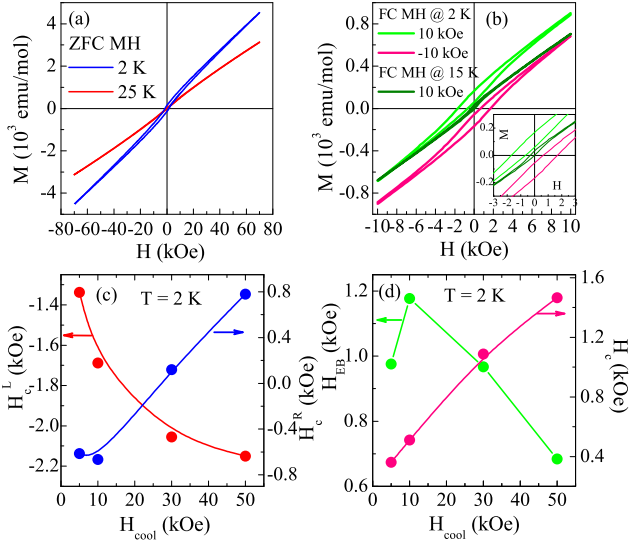}
	\caption{(a) shows the ZFC $M(H)$ at 2 and 25 K for Ba$_3$NiIr$_2$O$_9$. (b) The FC $M(H)$ at 2 K with $H_{cool}$ = 10 and -10 kOe and at 15 K with $H_{cool}$ = 10 kOe field are shown. Inset presents an expanded view of main panel plot (same unit) close to origin showing a shifting along field axis. (c) The left and right panel show the left and right coercive fields $H_c^L$ and $H_c^R$ with $H_{cool}$, respectively. (d) shows the EB field $H_{EB}$ (left panel) and coercive field $H_c$ (right panel) with $H_{cool}$ (see text).}
	\label{fig:Fig3}
\end{figure}

Magnetic field dependent moment $M(H)$ at 2 and 25 K are plotted in Fig. 3(a) which show a continuous linear increase with a coercive field ($H_c$) around 1410 and 330 Oe, respectively. A finite $H_c$ at 25 K suggests spin ordering even above $T_p$ where its substantial increase at 2 K is due to spin freezing and low thermal energy. The $M(H)$ data have also been collected after cooling the sample in a field $H_{cool}$ from high temperature, and then FC $M(H)$ loop are recorded with cycling the field from +$H_{cool}$ to -$H_{cool}$ and back to again +$H_{cool}$. We have used $H_{cool}$ = 5, 10, -10, 30 and 50 kOe. Fig. 3(b) depicts the FC $M(H)$ at 2 K with $H_{cool}$ = 10 and -10 kOe where both the data show closed loops with an opposite and symmetric shifting along the field axis. The variation of left and right coercive fields ($H_c^L$ and $H_c^R$, respectively) are shown in Fig. 3(c) with $H_{cool}$ at 2 K. An asymmetry in $M(H)$ ($H_c^L$ $\neq$ $H_c^R$) is usually caused by an exchange bias (EB) effect when a system with FM-AFM or even FM-SG interface is cooled in field. The external field biases the exchange interaction at interface and creates an unidirectional anisotropy, as a result a shifted FC $M(H)$ is observed.\cite{Nogue} We have calculated the quantifying exchange bias field $H_{EB}$ = $\left(|H_c^L + H_c^R|\right)$/2 and coercive field $H_{c}$ = $\left(|H_c^L - H_c^R|\right)$/2. At 2 K, we find $H_{EB}$ = 1175 Oe and $H_c$ = 506 Oe for $H_{cool}$ = 10 kOe and $H_{EB}$ = 1177 Oe and $H_c$ = 516 Oe for $H_{cool}$ = -10 kOe.

The closing of FC $M(H)$ as well as its opposite shifting with sign of $H_{cool}$ confirms a EB effect in present BNIO. The calculated $H_{EB}$ and $H_c$ with $H_{cool}$ are shown in Fig. 3(d). While $H_c$ shows a continuous increase, the $H_{EB}$ shows maximum around $H_{cool}$ = 1 kOe. Evolution of $H_{EB}$ with $H_{cool}$ is clear in Fig. 3(c) which shows though $|H_c^L|$ increases continuously the $H_c^R$ changes sign to positive at field $H_{cool}$ = 30 kOe and above. It is remarkable that in low $H_{cool}$, both $H_c^L$ and $H_c^R$ turns out to be negative which imply a strong exchange bias at interface where a reversal of moment even does not require a field reversal. Due to Zeeman magnetic coupling, the field cooling in high $H_{cool}$ promotes FM component significantly which results in positive $H_c^R$, thus giving a decrease in $H_{EB}$. The origin of EB effect in present BNIO is due to competition between intralayer AFM and interlayer FM interaction where a finite $H_{EB}$ $\sim$ 145 Oe at 15 K with $H_{cool}$ = 10 kOe (Fig. 3(b)) indicates an existence of both interaction above $T_p$. The observed $H_{EB}$ appears to be reasonably high considering present BNIO is a natural system unlike artificially designed interface in films and multilayers.

\begin{figure}
	\centering
		\includegraphics[width=7.0cm]{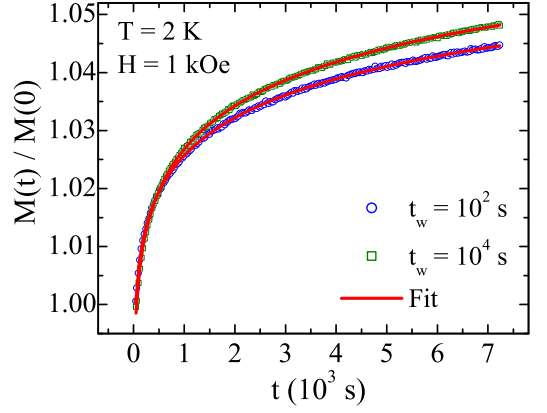}
	\caption{The normalized magnetic relaxation data $M(t)$ are shown with time $t$ for different wait time $t_w$. The lines are due to fitting with stretched exponential function (Eq. 1).}
	\label{fig:Fig4}
\end{figure}

Now turning our focus to investigate the SG behavior in BNIO, we have measured the time dependent magnetic relaxation $M(t)$ and aging effect which are very useful to understand the nonequilibrium ground state in SG.\cite{Binder} For $M(t)$ measurements, first the sample is cooled in zero field to 2 K, then after a wait time $t_w$ a magnetic field of 1 kOe field has been applied and data are collected. The normalized $M(t)$/$M(0)$ are plotted in Fig. 4 for $t_w$ = 10$^2$ and 10$^4$ s which show that $M(t)$ increases continuously without saturation even after 2 h. Further, $M(t)$ relaxation is found to be sensitive to wait time $t_w$ where the system waited for longer time at 2 K before field is applied (i.e., $t_w$ is higher), relaxes more. This is called aging effect. Both these magnetic relaxation and aging behavior signify the nonequilibrium ground state which is typical for SG dynamics.\cite{Binder} The $M(t)$ data can be fitted with standard stretched exponential function,\cite{Chamberlin}

\begin{eqnarray}
	M(t) = M_0 -M_r \exp\left[-\left(\frac{t}{\tau}\right)^{1-n}\right]
\end{eqnarray}

where $M_0$, $M_r$, $\tau$ and $n$ are the intrinsic magnetization, glassy component, characteristic time constant and exponent, respectively. Fig. 4 shows a good fitting with Eq. 1 where the obtained fitting parameters are given in Table II. It is evident in table that parameters $M_0$, $M_r$, $\tau$ vary with $t_w$ but the exponent $n$, which is the measure of interaction strength between the local moments, vary insignificantly. Nonetheless, these results suggest a nonequilibrium ground state in present BNIO which further supports SG behavior at low temperature. 

\begin{table}
\caption{\label{label} Parameters obtained from fitting of stretched exponential function (Eq. 1) with magnetic relaxation $M(t)$ as shown in Fig. 4.}
\begin{ruledtabular}
\begin{tabular}{ccccc}
t$_{w}$ &M$_{0}$ &M$_{r}$ &$\tau$ &n\\
(s) &(emu/mol) &(emu/mol) &(s)\\         
\hline
10$^{2}$ &102.45 &6.9(1) &2885(113) &0.63(1)\\
10$^{4}$ &96.92 &7.0(1) &2469(80) &0.61(1)\\
\end{tabular}
\end{ruledtabular}
\end{table}

The spin dynamics in BNIO has been probed with ac susceptibility ($\chi_{ac}$) which not only employs low measuring field but also allows to use different probe time to investigate the material by changing its measurement frequency ($f$). Fig. 5(a) shows the real part of first order $\chi_{ac}$ ($\chi_{ac}^{/}$) with temperature for BNIO at different frequencies. In agreement with dc $\chi_{ZFC}$ (Fig. 2(a)), the $\chi_{ac}^{/}$ exhibits peak around $T_p$ in Fig. 5(a). For instance, at lowest $f$ = 1.3 Hz the peak in $\chi_{ac}^{/}$ occurs at $T_f$ $\sim$ 8.58 K but with increasing $f$ both $\chi_{ac}^{/}$ decreases and its peak shifts to higher temperature which are the hallmark of glassy systems. \cite{Binder} The upper inset of Fig 5(a) shows the imaginary part $\chi_{ac}^{//}$ at representative $f$ = 133 Hz which shows peak at 8.4 K. However, the shifting of peak is not very significant, giving $\Delta T_f$ $\sim$ 0.22 K. We calculate the quantifying parameter, $\delta T_f$ (= $\Delta T_f$/$T_{f} \Delta \log f$) around 0.009. The $\delta T_f$ comes out very low though its value lies in the range of classical SG and interacting particle systems (0.005 - 0.05) though much lower than values (0.1 -0.3) for superparamagnetic systems.\cite{Anand,Binder,AKP,Ma,Krizan}

We have further analyzed the $T_f(f)$ following different phenomenological models. According to N$\acute{e}$el-Arrhennius law, the relaxation time $\tau$ [= $(2\pi f)^{-1}$] for noninteracting particle system varies with temperature as, $\tau$ = $\tau_{0} \exp\left[E_a/\left(k_B T\right)\right]$ where $\tau_0$ is the microscopic relaxation time, $E_a$ is the activation energy and $k_B$ is the Boltzmann constant. The straight line fitting of $\log \tau$ vs $T^{-1}$ for BNIO gives unphysical parameters $\tau_{0}$ = 8.4 $\times$ 10$^{-112}$ s and $E_a/k_B$ = 943.2 K, though the fitting looks reasonable (lower inset of Fig. 5(a)). The frequency dependent $T_f$ has further been analyzed with Vogel-Fulcher (VF) law, $\tau = \tau_{0} \exp \left[\frac{E_a}{k_{B} (T - T_0)}\right]$, where an inclusion of $T_0$ (usually $\leq$ $T_f$) takes care the spin-spin or interparticle interaction. The fitting of VF law (Fig. 5(b)) for present data yields $\tau_0$ = 1.9 $\times$ 10$^{-7}$ s, $E_a$/$k_B$ = 3.25 K and $T_0$ = 8.3 K. While both $E_a$/$k_B$ and $T_0$ show reasonable values but the $\tau_0$ comes out very high compared to SG systems. Nonetheless, the parameter ($T_f$ - $T_0$)/$T_f$, which has been used to characterize the SG, is calculated to be $\sim$ 0.03 for present BNIO showing a close agreement with both metallic as well as insulating SG systems.\cite{THOLENCE, Yeshurun} The $T_f(f)$ has further been analyzed with dynamic scaling hypothesis which describes that near the glass transition temperature $T_f$, the $\tau$ shows a relation with the spin-spin correlation length $\xi$ as, $\tau$ $\propto$ $\xi^z$. Since $\xi$ diverges on approaching $T_f$, it gives an empirical relation, $\tau = \tau_{0} \left(T/T_{f} - 1\right)^{-z\nu}$ where $\nu$ is the critical exponent related to $\xi$ and $z$ is the dynamical scaling exponent.\cite{Binder} The $\tau_0$ signifies here the relaxation time of individual spin. The Fig. 5(c) shows the best fitting of scaling law for BNIO with $\tau_{0}$ = 4.8 $\times$ 10$^{-12}$ s, $T_f$ = 8.48 K and $z\nu$ = 5.32. The exponent $z\nu$ has shown wide variation with the materials,\cite{Souletie,Yeshurun,Anand} however, our obtained $z\nu$ is close to the predicted value (4) for 3D Ising model \cite{Binder1} and also to the experimentally obtained value (5.5) for canonical SG system CuMn$_{4.6\%}$.\cite{Souletie} The obtained $\tau_0$ nicely matches with the values for SG systems (10$^{-11}$ - 10$^{-12}$ s). The validity of dynamical scaling analysis in BNIO shows critical slowing down of relaxation time which again confirms the SG like low -$T$ behavior.

\begin{figure}
	\centering
		\includegraphics[width=8.5cm]{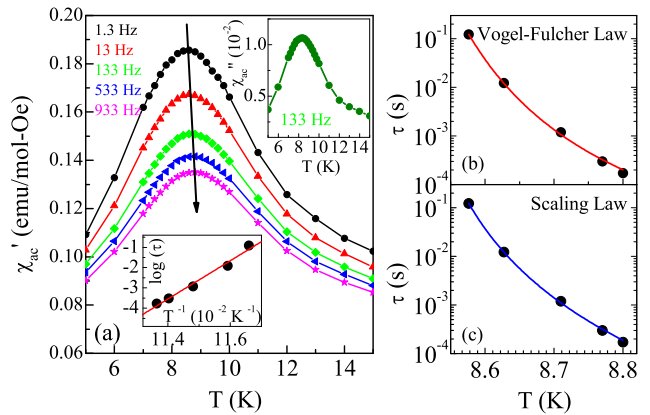}
	\caption{(a) shows the real part of 1st order ac susceptibility $\chi_{ac}^/$ with temperature at different frequencies for Ba$_3$NiIr$_2$O$_9$. The upper inset shows corresponding imaginary part at $f$ = 133 Hz while the lower inset shows fitting following N$\acute{e}$el-Arrhennius law. (b) and (c) show the best fitting of relaxation time $\tau$ following Vogel-Fulcher and scaling law, respectively.}
	\label{fig:Fig5}
\end{figure}

To understand the thermodynamic behavior in present BNIO, temperature dependent specific heat $C_p(T)$ has been measured down to 2 K and in applied magnetic fields of 0, 60 and 120 kOe (Fig. 6(a)). An expanded plot of $C_p(T)$, in left inset of Fig. 6(a), shows an anomaly or broad hump around $T_f$ $\sim$ 8.5 K. The hump in $C_p(T)$ is suppressed with magnetic field though its trace still exists in field as high as 120 kOe. This robust and broad hump in $C_p(T)$ indicates a low temperature glassy behavior. Similar to $\chi_{ZFC}$, the shoulder in $C_p(T)$ hump moves to lower temperature with increasing field. The anomaly in $C_p(T)$ is also observed around $\theta_{CW}$ ($\sim$ 28.7 K) which is evident in its derivative $dC_p/dT$, as shown in right inset of Fig. 6(a). The magnetic contribution $C_m$ to the specific heat is estimated by subtracting its lattice contribution $C_l$ from $C_p$. In absence of suitable nonmagnetic reference sample, the lattice specific heat $C_l$ has been determined by fitting the zero-field $C_p(T)$ data in high temperature regime (30 - 156 K) to a combined one Debye and three Einstein (1D + 3E) functions of lattice specific heat. The so obtained $C_l(T)$ then extrapolated to entire temperature range down to 2 K, shown as continuous solid line in Fig. 6(a). We have used following Debye (Eq. 2) and Einstein (Eq. 3) functions to fit the experimentally observed $C_p(T)$ data. The $C_D$ and $C_{Ei}$ in Eqs. 2 and 3 are the weightage factors corresponding to an acoustic and optical modes of atomic vibrations and $\theta_D$ and $\theta_{Ei}$ are the Debye and Einstein temperatures, respectively.
   
\begin{eqnarray}
	C_{Debye}(T) = C_{D}\left[9R\left(\frac{T}{\theta_{D}}\right)^3\int_{0}^{\frac{\theta_{D}}{T}}{\frac{x^4e^x}{(e^x-1)^2}}\right]
\end{eqnarray}

\begin{eqnarray}
	C_{Einstein}(T) = \sum_{i} C_{E_i}\left[3R\left(\frac{\theta_{E_i}}{T}\right)^2\frac{e^{(\theta_{E_i}/T)}}{\left(e^{(\theta_{E_i}/T)}-1\right)^2}\right]
\end{eqnarray}

From fitting, we obtain the weightage factors $C_D$ : $C_1$ : $C_2$ : $C_3$ = 1 : 4 : 5 : 5, where the total sum $C_D$ + $\sum_i C_{Ei}$ matches with the total number of atoms (15) per formula unit in BNIO sample. We further obtain the Debye temperature $\theta_D$ = 103.2 K,and Einstein temperatures $\theta_{E_1}$ = 150.1 K, $\theta_{E_2}$ = 684.5 K and $\theta_{E_3}$ = 314 K.

The so obtained magnetic specific heat $C_m$ is shown in Fig. 6(b). Unlike a $\lambda$-like peak in case of magnetic transitions, the $C_m/T$ in Fig. 6(b) shows a broad hump across $T_f$. In zero field, the $C_m/T$ emerges below $\theta_{CW}$ and shows peak around 5.5 K. With increasing field, the peak shifts slightly to low temperature (in 60 and 120 kOe field the peak occurs around 5.3 and 5.0 K, respectively) and its value decreases. In case of long-range magnetic ordering, the peak in $C_m/T$ should suppress completely by a field of $k_B T_f$/$g \mu_B$ ($\sim$ 60 kOe) but that is not the case here in Fig. 6(b). A robust broad hump in $C_m/T$ and its smearing out to higher temperature in high field are the indicative of SG behavior.\cite{Raju} The magnetic entropy $S_m$, which has been calculated by integrating $C_m/T$ with $T$, are shown in right panel of Fig. 6(b). At temperature above $\sim$ 20 K, the $S_m$ almost saturates at $\sim$ 4.93 J/mol-K which is significantly low and only about 54\% of the expected value 9.14 J/mol-K [$R \ln(2S + 1)$, $R$ is the gas constant] for 100\% spin freezing with $S$ = 1 system. In presence of field, the $S_m$ further lowers but the change is monotonic with the applied field (Fig. 6(b)). This low $S_m$ indicates a highly degenerate low-energy ground state which is typical for disordered and glassy systems.
 
\begin{figure}
	\centering
		\includegraphics[width=8.5cm]{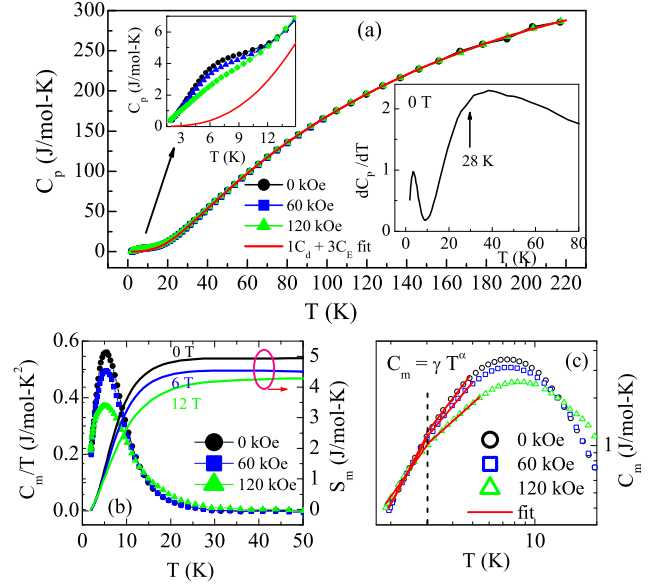}
	\caption{(a) The main panel shows the temperature dependent specific heat $C_p$ at different fields along with lattice specific heat (solid red line) obtained from fitting with combined Debye and Einstein model for Ba$_3$NiIr$_2$O$_9$ (see text). The left and right inset shows an expanded view and derivative $dC_p/dT$, respectively. (b) shows the temperature dependent magnetic specific heat $C_m/T$ in different magnetic fields (left panel) and the calculated magnetic entropy $S_m$ (right panel). (c) The temperature- and field-dependent $C_m$ has been shown along with fitting ($C_m$ = $\gamma T^{\alpha}$). The vertical dotted line marks the different fitting region.}
	\label{fig:Fig6}
\end{figure}

The temperature variation of $C_m$ can be described with $C_m$ = $\gamma T^{\alpha}$ power-law behavior. The $\log-\log$ plotting in Fig. 6(c) clearly shows two distinct linear regimes below $T_f$. In zero field, we obtain values $\alpha$, $\gamma$ = 1.72(6), 184(16) mJ/mol-K$^{2.7}$ between 3 - 5 K and 2.49(2), 75.4(1) mJ/mol-K$^{3.5}$ in range 2 - 3 K. In 60 kOe field, the corresponding values are $\alpha$, $\gamma$ = 1.56(5), 205(14) mJ/mol-K$^{2.5}$ and 2.39(5), 80.9(4) mJ/mol-K$^{3.5}$ and in 120 kOe the values are 1.18(3), 280(11) mJ/mol-K$^{2}$ and 1.85(5), 131.2(7) mJ/mol-K$^{3}$, respectively. In other word, with increasing field, the $\gamma$ increases while $\alpha$ decreases in both temperature range. Nonetheless, the $\gamma$ turns out quite high even in zero field. Usually, a linear temperature dependence, $C_m \propto T$, is observed in case of canonical SG systems,\cite{Binder} however, exceptions are also documented in literature based on the material properties. For instance, a $T^{1.5}$ dependence in many SG systems such as, diluted alloys,\cite{Binder} pyrochlore oxide Yb$_2$Mn$_2$O$_7$,\cite{Greedan} hexagonal compound Na$_{0.70}$MnO$_2$,\cite{Luo} amorphous Er$_x$Ni$_{100-x}$,\cite{Thomson} etc. has been explained with spin-wave like excitations in FM components. Further, a quadratic $T^2$ dependance in pyrochlore NaCaNi$_2$F$_7$ SG has been ascribed to spin-wave excitations in quasi 2D AFM ordering.\cite{Krizan} Even, a $T^{2.5}$ dependence has been shown in spin-1/2 TLAF LiNiO$_2$ due to an orbital degeneracy.\cite{Kitaoka} In layered Ba$_3$NiSb$_2$O$_9$ which has an 2D AFM ordering with $T_N$ = 13.5 K, a $T^3$ dependance below $T_N$ has been ascribed due to 3D magnon scattering.\cite{Cheng}

In present BNIO, the $C_m$ = $\gamma T^{\alpha}$ dependence with exponent $\alpha$ = 1.72(6) (3 - 5 K) and 2.49(2) (2 - 3 K) and its evolution with magnetic field is quite intriguing. Given that a competition between strong intralayer AFM and weak interlayer FM interaction (via Ir$^{5+}$) is believed to cause the SG-like dynamics, both AFM- and FM-type spin-wave excitations likely to contribute in $\alpha$, giving it an intermediate value (1.72). At low temperature (2 - 3 K), however, with the hardening of SG state the $\alpha$ increases to value $\sim$ 2.5. In applied field, due to Zeeman coupling the FM ordering is favored which is believed to lower the $\alpha$ but its value remains still above 1.

\begin{figure}
	\centering
		\includegraphics[width=7cm]{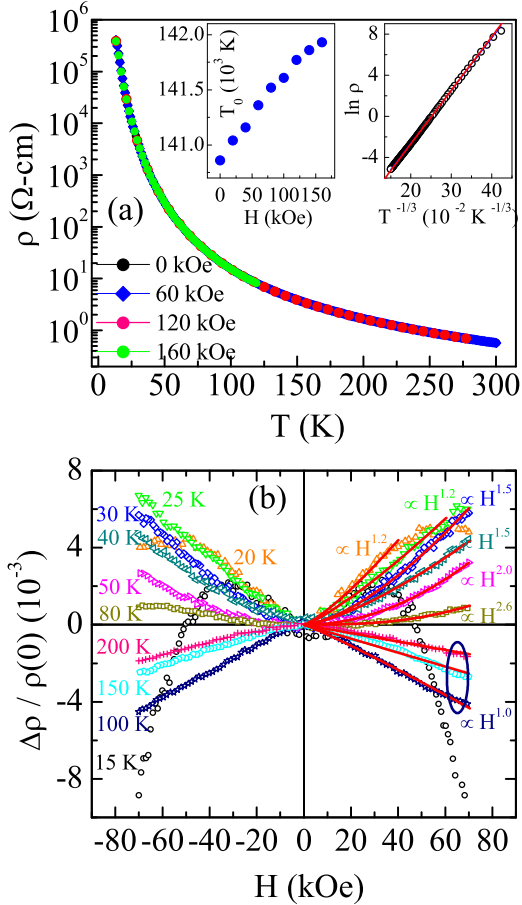}
	\caption{(a) shows temperature variation of resistivity $\rho$ for Ba$_3$NiIr$_2$O$_9$ in different magnetic fields. Right inset shows linear fitting of $\rho(T)$ data in zero field with Mott's 2D VRH model and left inset shows the variation of parameter $T_0$ with magnetic field. (b) Magnetoresistance (MR) collected at different temperatures are shown against magnetic field, showing a both temperature- and field-dependent sign change. The solid lines are due to fitting with, MR $\propto H^q$.}
	\label{fig:Fig7}
\end{figure}

The temperature dependent electrical resistivity $\rho(T)$ are shown in Fig. 7(a) for different magnetic fields. As evident in figure, the insulating state in BNIO is very robust against magnetic field. The $\rho(T)$ increases roughly by six orders at low temperature where below $\sim$ 13 K the $\rho(T)$ goes beyond the measurement capacity. This thermally activated charge conduction can be explained well with Mott's 2D variable-range-hopping (VRH) model \cite{N}, $\rho$ = $\rho_0 \exp\left[\left(T_0/T\right)^{1/3}\right]$ where $T_0$ is the characteristic temperature shown as, 21.2/$\left[k_B N(\epsilon_F) \xi^3\right]$ with $k_B$ and is the Boltzmann constant, $N(\epsilon_F)$ is the density of states at Fermi level and $\xi$ is the localization length. The right inset of Fig. 7(a) shows the linear plotting of $\ln \rho$ vs $T^{-1/3}$ in zero field over the temperature range. In magnetic field, the charge transport also follows 2D VRH model (not shown) where the parameter $T_0$ increases with field (left inset of Fig. 7(a)). The BNIO being a hard insulator, the change in $T_0$ with field is unlikely contributed by $N(\epsilon_F)$, rather a decrease in $\xi$ with field explains this change well.

The magnetoresistance (MR), calculated as $\Delta \rho$/$\rho(0)$ = [$\rho(H)$ - $\rho(0)$]/$\rho(0)$, are shown in Fig. 7(b) across the temperature down to 15 K and field up to $\pm$70 kOe. The MR is symmetric with the sign of fields. While above 100 K the MR is negative showing a linear field dependence ($\propto H$), but it changes to positive value below $\sim$ 80 K and  shows a $\propto H^q$ dependence. The fitting of MR data (shown only for positive $H$) in Fig. 7(b) shows the exponent $q$ gradually decreases with lowering the temperature. Note, that this crossover temperature ($\sim$ 80 K) is close to the onset of irreversibility in $\chi(T)$ (Fig. 2(a)), indicating that the spin ordering has dominant role on change transport. A field induced crossover from positive to negative MR has further been evidenced at 15 K though a precursor is seen at 20 K. While this temperature- and field-dependent crossover in MR is intriguing, it can be noted that present material has dominant SOC from Ir in addition to magnetic ordering of respective Ni/Ir sublattices. We believe that an onset of spin correlation switches the positive MR around 80 K while its softening with decreasing temperature (exponent $q$ decreases) and field induced crossover at low $T$ are likely due to an appearance of Ni layer AFM-ordering where the applied field dominates over the moment and similar negative MR is realized.  This temperature- and field-dependent MR sheds light to understand the nature of magnetism in this material. These results bring out an exotic interplay among magnetic moment, field and SOC on the charge transport in BNIO where the same has been demonstrated in another SOC dominant material Sr$_2$IrO$_4$.\cite{Bhatti1}

In conclusion, using detailed dc and ac magnetization and specific heat measurements we have shown low temperature spin-glass behavior ($T_f$ $\sim$ 8.5 K) in Ba$_3$NiIr$_2$O$_9$. This material has layer structure of Ni$^{2+}$ (spin-1) spins engaging in triangular-lattice antiferromagnetic interaction while layers are magnetically connected via Ir$^{5+}$. Along with large exchange bias, this material shows magnetic relaxation and aging effect which characterizes the glassy behavior. The magnetic specific heat indicates a highly degenerate ground state and shows a $C_m = \gamma T^\alpha$ dependence where temperature and magnetic field has influence on both $\gamma$ and $\alpha$. An insulating state prevails over the temperature, though a change in sign of magnetoresistance has been observed which is due to an interplay between SOC and magnetic moment.

\section{Acknowledgment}
We acknowledge UGC-DAE CSR, Indore for the magnetization, specific heat and electrical transport measurements. We are thankful to Alok Banerjee for the magnetization measurements and discussion. We thank DST and DST-PURSE, India for the financial support. S.G. is thankful to UGC, India for financial support.


\begin{thebibliography}{}
\bibitem{Ramirez} A. P. Ramirez, Annu. Rev. Mater. Sci. \textbf{24}, 453 (1994).
\bibitem{Nakatsuji1} S. Nakatsuji, K. Kuga, K. Kimura, R. Satake, N. Katayama, E. Nishibori, H. Sawa, R. Ishii, M. Hagiwara, F. Bridges, T. U. Ito, W. Higemoto, Y. Karaki, M. Halim, A. A. Nugroho, J. A. Rodriguez-Rivera, M. A. Green, C. Broholm, Science \textbf{336}, 559 (2012).
\bibitem{Zhou} H. D. Zhou, E. S. Choi, G. Li, L. Balicas, C. R. Wiebe, Y. Qiu, J. R. D. Copley, and J. S. Gardner, Phys. Rev. Lett. \textbf{106}, 147204 (2011).
\bibitem{Cheng} J. G. Cheng, G. Li, L. Balicas, J. S. Zhou, J. B. Goodenough, Cenke Xu, and H. D. Zhou, Phys. Rev. Lett.  \textbf{107}, 197204 (2011).
\bibitem{Okamoto} Y. Okamoto, M. Nohara, H. Aruga-Katori, and H. Takagi, Phys. Rev. Lett. \textbf{99}, 137207 (2007).
\bibitem{Nag} A. Nag, S. Middey, S. Bhowal, S. K. Panda, R. Mathieu, J. C. Orain, F. Bert, P. Mendels, P. G. Freeman, M. Mansson, H. M. Ronnow, M. Telling, P. K. Biswas, D. Sheptyakov,S. D. Kaushik, V. Siruguri, C. Meneghini, D. D. Sarma, I. Dasgupta, and S. Ray, Phys. Rev. Lett. \textbf{116}, 097205 (2016).
\bibitem{Nakatsuji} S. Nakatsuji, Y. Nambu, H. Tonomura, O. Sakai, S. Jonas, C. Broholm, H. Tsunetsugu, Y. Qiu, Y. Maeno, Science \textbf{309}, 5741 (2005).
\bibitem{MacLaughlin} D. E. MacLaughlin, Y. Nambu, S. Nakatsuji, R. H. Heffner, Lei Shu, O. O. Bernal, and K. Ishida, Phys. Rev. B \textbf{78}, 220403 (2008).
\bibitem{Krizan} J. W. Krizan and R. J. Cava, Phys. Rev. B \textbf{92}, 014406 (2015). 
\bibitem{Fu} Z. Fu, Y. Zheng, Y. Xiao, S. Bedanta, A. Senyshyn, G. G. Simeoni, Yixi Su, U. Rucker, P. Kogerler, and T. Bruckel, Phys. Rev. B \textbf{87}, 214406 (2013).
\bibitem{Greedan}  J. E. Greedan, N. P. Raju, A. Maignan, Ch. Simon, J. S. Pedersen, A. M. Niraimathi, E. Gmelin, and M. A. Subramanian, Phys. Rev. B \textbf{54}, 7189 (1996).
\bibitem{Raju} N. P. Raju, E. Gmelin, and R. K. Kremer, Phys. Rev. B \textbf{46}, 9 (1992).
\bibitem{Ma} Z. Ma, J. Wang, Z.-Y. Dong, J. Zhang, S. Li, S.-H. Zheng, Y. Yu, W. Wang, L. Che, K. Ran, S. Bao, Z. Cai, P. Čermák, A. Schneidewind, S. Yano, J. S. Gardner, X. Lu, S.-L Yu, J.-M. Liu, S. Li, J.-Xin Li, and J. Wen, Phys. Rev. Lett. \textbf{120}, 087201 (2018).
\bibitem{Gaudet} J. Gaudet, E. M. Smith, J. Dudemaine, J. Beare, C. R. C. Buhariwalla, N. P. Butch, M. B. Stone, A. I. Kolesnikov, Guangyong Xu, D. R. Yahne, K. A. Ross, C. A. Marjerrison, J. D. Garrett, G. M. Luke, A. D. Bianchi, and B. D. Gaulin, Phys. Rev. Lett. \textbf{122}, 187201 (2019).
\bibitem{Anderson} P. W. Anderson, Phys. Rev. \textbf{102}, 1008 (1956).
\bibitem{Villain} J. Villain, Z. Physik B \textbf{33}, 31 (1979).
\bibitem{Kim} B. J. Kim, H. Jin, S. J. Moon, J. Y. Kim, B.-G. Park, C. S. Leem, J. Yu, T. W. Noh, C. Kim, S. J. Oh, J. H. Park, V. Durairaj, G.Cao, and E. Rotenberg, Phys. Rev. Lett. \textbf{101}, 076402 (2008).
\bibitem{Kim1} B. J. Kim, H. Ohsumi, T. Komesu, S. Sakai, T. Morita, H.Takagi, and T. Arima, Science \textbf{323}, 1329 (2009).
\bibitem{Bremholm} M. Bremholm, S. E. Dutton, P. W. Stephens, and R. J. Cava, J. Solid State Chem. \textbf{184}, 601 (2011).
\bibitem{Bhowal} S. Bhowal, S. Baidya, I. Dasgupta, T. S.-Dasgupta, Phys. Rev. B \textbf{92}, 121113 (2015).
\bibitem{Kharkwal} K. C. Kharkwal, R. Roy, H. Kumar, A. K. Bera, S. M. Yusuf, A. K. Shukla, K. Kumar, S. Kanungo, and A. K. Pramanik, Phys. Rev. B \textbf{102}, 174401 (2020).
\bibitem{Cao} G. Cao, T. F. Qi, L. Li, J. Terzic, S. J. Yuan, L. E. DeLong, G. Murthy, and R. K. Kaul, Phys. Rev. Lett. \textbf{112}, 056402 (2014).
\bibitem{Khan} Md S. Khan, A. Bandyopadhyay, A. Nag, V. Kumar, A. V. Mahajan, and S. Ray, Phys. Rev. B \textbf{100}, 064423 (2019).
\bibitem{Kundu} S. Kundu, A. Shahee, A. Chakraborty, K. M. Ranjith, B. Koo, J. Sichelschmidt, M. T. F. Telling, P. K. Biswas, M. Baenitz, I. Dasgupta, S. Pujari, and A. V. Mahajan, Phys. Rev. Lett. \textbf{125}, 267202 (2020).
\bibitem{Lu} Z. Lu, L. Ge, G. Wang, M. Russina, G. Günther, C. R. dela Cruz, R. Sinclair, H. D. Zhou, and J. Ma, Phys. Rev. B \textbf{98}, 094412 (2018).
\bibitem{Hwang} J. Hwang, E. S. Choi, F. Ye, C. R. Dela Cruz, Y. Xin, H. D. Zhou, and P. Schlottmann, Phys. Rev. Lett. \textbf{109}, 257205 (2012).
\bibitem{suite} FULLPROF suite, http://www.ill.eu/sites/fullprof/.
\bibitem{Bhatti} I. N. Bhatti, R. S. Dhaka, and A. K. Pramanik, Phys. Rev. B. \textbf{96}, 144433 (2017).
\bibitem{Harish} H. Kumar, R. S. Dhaka, and A. K. Pramanik, Phys. Rev. B. \textbf{95}, 054415 (2017).
\bibitem{Zhu} W. K. Zhu, M. Wang, B. Seradjeh, F. Yang, and S. X. Zhang, Phys. Rev. B \textbf{90}, 054419 (2014).
\bibitem{Liu} X. Liu, Y. Cao, B. Pal, S. Middey, M. Kareev, Y. Choi, P. Shafer, D. Haskel, E. Arenholz, 
and J. Chakhalian, Phys. Rev. M \textbf{1}, 075004 (2017).
\bibitem{Babita} B. Baruwati, R. K. Rana, and S. V. Manorama, J. of Appl. Phys. \textbf{101}, 014302 (2007).
\bibitem{Elipe} A. R. Gonzalez-Elipe, J. P. Holgado, R. Alvarez, and G. Munuera, J. Phys. Chem. \textbf{96}, 3080 (1992).
\bibitem{Peck} M. A. Peck and M. A. Langell, Chem. Mater. \textbf{24}, 4483 (2012).
\bibitem{Luo} L. B. Luo, Y. G. Zhao, G. M. Zhang, S. M. Guo, Z. Li and J. L. Luo, Phys. Rev. B \textbf{75}, 125115 (2007).
\bibitem{Anand} V. K. Anand, D. T. Adroja, and A. D. Hillier, Phys. Rev. B \textbf{85}, 014418 (2012).
\bibitem{Kitaoka} Y. Kitaoka, T. Kobayashi, A. Koda, H. Wakabayashi, Y. Niino, H. Yamakage, S. Taguchi, K. Amaya, K. Yamaura, M. Takano, A. Hirano and R. Kanno, J. Phys. Soc. Jpn. \textbf{67}, 11 (1998). 
\bibitem{Carlin}  R. L. Carlin, Magnetochemistry (Springer, Berlin, 1986), Chap. 4.
\bibitem{Shirata} Y. Shirata, H. Tanaka, T. Ono, A. Matsuo, K. Kindo, and H. Nakano, J. Phys. Soc. Jpn. \textbf{80}, 093702 (2011). 
\bibitem{Nogue} J. Nogu$\acute{e}$s and I. K. Schuller, J. of Magn. Magn. Mater. \textbf{192}, 203 (1999).
\bibitem{Chamberlin} R. V. Chamberlin, G. Mozurkewich, and R. Orbach, Phys. Rev. Lett. \textbf{52}, 867 (1984).
\bibitem{Binder} K. Binder and A. P. Young, Rev. Mod. Phys. \textbf{58}, 4 (1986).
\bibitem{AKP} A. K. Pramanik and A. Banerjee, Phys. Rev. B \textbf{82}, 094402 (2010).
\bibitem{THOLENCE} J.-L. Tholence, Physica \textbf{126B}, 157 (1984).
\bibitem{Yeshurun} Y. Yeshurun, J. L. Tholence, J. K. Kjems and B. Wanklyn, J. Phys. C: Solid State Phys. \textbf{18}, L483 (1985).
\bibitem{Souletie} J. Souletie and J. L. Tholence, Phys. Rev. B \textbf{32}, 1 (1985). 
\bibitem{Binder1} K. Binder and A. P. Young, Phys. Rev. B \textbf{29}, 5 (1984).
\bibitem{Thomson} J. O. Thomson and J. R. Thomson, J. Phys. F: Met. Phys. \textbf{11}, 247 (1981).
\bibitem{N} Mott N 1993 Conduction in Non-Crystalline Materials (Oxford: Clarendon Press).
\bibitem{Bhatti1} I. N. Bhatti and A.K. Pramanik, J. of Magn. Magn. Mater. \textbf{422}, 141 (2017). 
\end{thebibliography}
\end{document}